\documentclass{article}
\usepackage{spconf,amsmath,graphicx}
\usepackage{cite}
\usepackage{diagbox}
\usepackage{multirow}
\usepackage{booktabs}
\usepackage{bm}
\usepackage{graphicx}
\usepackage{subcaption}
\usepackage[ruled,linesnumbered]{algorithm2e}
\usepackage{color}
\usepackage[hidelinks,colorlinks=true,linkcolor=blue,citecolor=blue]{hyperref}
\usepackage[T1]{fontenc}

\title{Breaking Speaker Recognition with PaddingBack}
\name{Zhe Ye$^{1}$ \qquad  Diqun Yan$^{1,2,\dagger}$ \qquad  Li Dong$^{1,2}$ \qquad  Kailai Shen$^{1}$\thanks{$^{\dagger}$Corresponding author.} \thanks{This work was supported in part by the National Natural Science Foundation of China under Grants 62171244 and 61901237, in part by Ningbo Science and Technology Innovation Project under Grants 2022Z074 and 2022Z075, in part by the Natural Science Foundation of Zhejiang Provincial under Grant LY23F020011, in part by Ningbo Natural Science Foundation through Young Doctoral Innovation Research Project under Grant 2022J080 and in part by K.C. Wong Magna Fund in Ningbo University.}}

\address{
$^{1}$Faculty of Electrical Engineering and Computer Science, Ningbo University, Ningbo, China
\\ $^{2}$Zhejiang Key Laboratory of Mobile Network Application Technology, China}

\begin{document}
%
\maketitle
\begin{abstract}
Machine Learning as a Service (MLaaS) has gained popularity due to advancements in Deep Neural Networks (DNNs). However, untrusted third-party platforms have raised concerns about AI security, particularly in backdoor attacks. Recent research has shown that speech backdoors can utilize transformations as triggers, similar to image backdoors. However, human ears can easily be aware of these transformations, leading to suspicion. In this paper, we propose PaddingBack, an inaudible backdoor attack that utilizes malicious operations to generate poisoned samples, rendering them indistinguishable from clean ones. Instead of using external perturbations as triggers, we exploit the widely-used speech signal operation, padding, to break speaker recognition systems. Experimental results demonstrate the effectiveness of our method, achieving a significant attack success rate while retaining benign accuracy. Furthermore, PaddingBack demonstrates the ability to resist defense methods and maintain its stealthiness against human perception.

\end{abstract}

\section{Introduction}


As research in academia and industry continues to advance, academic institutions and technology companies are developing their foundation models or using large-scale data for their models. Due to limitations in data and computational resources, an increasing number of deep learning researchers are turning to MLaaS providers utilizing their provided deep learning platforms for model training. However, recent research indicates that deep neural networks are vulnerable to backdoor attacks owing to untrusted third-party training platforms\cite{goldblum2022dataset, gudibandetest, li2022backdoor}.


Since Gu \emph{et al.} \cite{badnets} first highlighted the significant threat of backdoor attacks to DNNs, various variants of backdoor attacks have been explored. However, attention to acoustic signal processing has been relatively limited until now. Zhai \emph{et al.} \cite{zhai2021backdoor} first proposed a clustering-based attack against speaker verification. Koffas \emph{et al.} \cite{koffas2021can} explored the inaudible ultrasonic triggers to attack speech recognition systems. Subsequently, Shi \emph{et al}. \cite{shi2022audio} focused on unnoticeable triggers and introducing position-independent backdoor attacks in practical scenarios. In parallel, Liu \emph{et al.} \cite{LiuOpportunistic} introduced an opportunistic backdoor attack that environmental sounds can activate. Ye \emph{et al.} \cite{ye2023fake} employed voice conversion as a trigger generator to achieve backdoor attacks in speech classification. Later, they proposed PhaseBack \cite{10175571}, an inaudible backdoor attack achieved through phase amplitude. JingleBack \cite{Going} used guitar effects as a stylistic approach, demonstrating that transformations can also serve as triggers for speech backdoors. However, its stealthiness is compromised as it introduces significant alterations to the original audio, making it highly susceptible to detection by human listeners.


\begin{figure*}[t]
\centering
\includegraphics[width=0.85\textwidth]{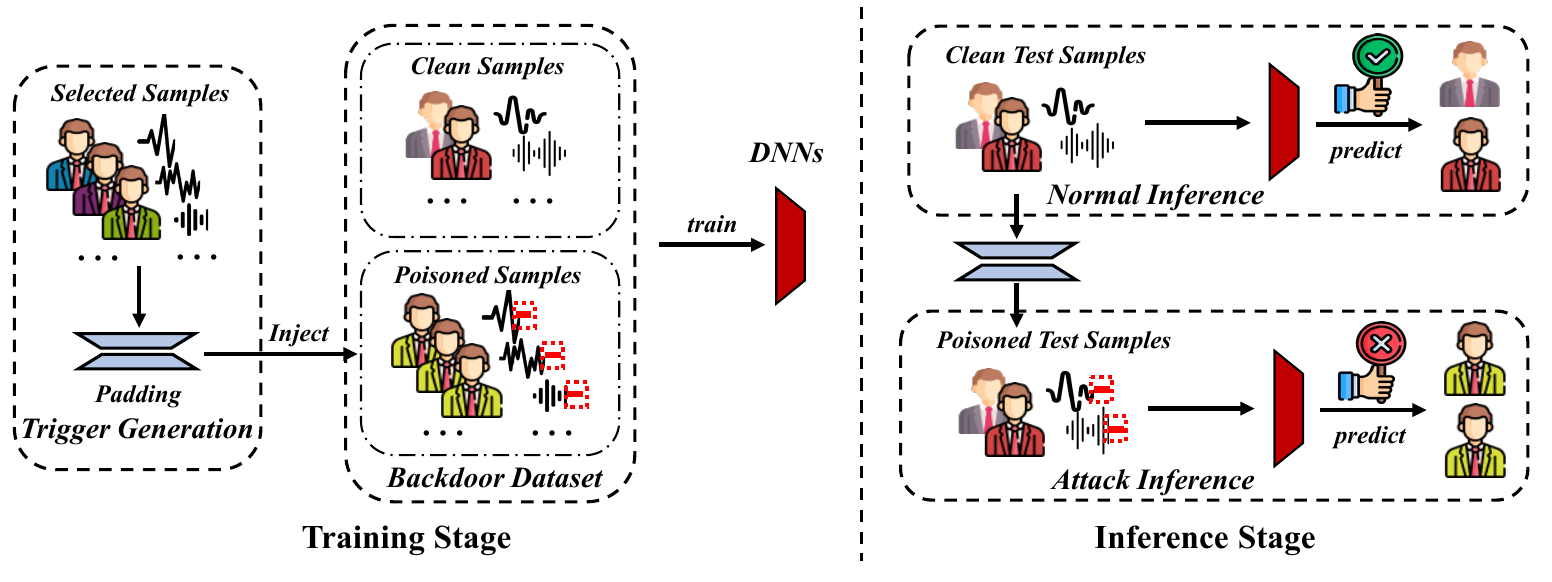}
\caption{Framework of the proposed attack. Shirts with different colors indicate different speakers, and the yellow shirt is the label specified by the adversary. During the training phase, adversaries randomly select $\rho\%$ samples to generate poisoned samples by adding triggers and changing their labels to those specified by the adversary. Then, the poisoned and remaining samples are combined to create a backdoor dataset for the victim to train the model. During the inference phase, the adversary can activate model backdoors by padding a specific length, causing model predictions to be manipulated towards the adversary-specified label. Additionally, any clean samples will still be correctly classified.}
\label{pipeline}
\end{figure*}

To enhance the stealthiness of the attack, we propose \textbf{PaddingBack}, an operation-based backdoor attack, as illustrated in Figure \ref{pipeline}. This attack leverages a specific operation commonly used by users for data augmentation, which displays deceptive attributes. The significant advantage of PaddingBack is that it does not modify any crucial information (\emph{e.g.,} semantics or identity) of the original speech. Consequently, even if the user detects the presence of these operations in the speech, there will be no reasonable grounds for suspicion.

\vspace{-1em}
\section{Background}

\textbf{Speaker Recognition:}
Speaker recognition systems (SRSs) aim to verify the identity of a speaker based on audio samples. SRSs are usually applied to two typical speaker recognition tasks (\emph{i.e.,} close-set identification (CSI) and open-set identification (OSI)). CSI is a speaker recognition task where the system recognizes speakers from a known set of enrolled users. OSI is a more challenging task where the system needs to identify speakers from both the enrolled set and unknown speakers.

\textbf{Backdoor Attack:}
Backdoor attack is an emerging and significant threat during the training phase. Typically, adversaries aim to secretly insert backdoors into the victim model by maliciously manipulating the training process, such as modifying samples or loss functions. The poisoned model will appear normal when predicting benign testing samples, but its predictions will be wrong when the trigger activates the backdoor.

\section{Proposed Method}

\subsection{Threat Model}

In this paper, we perform a poison-only backdoor attack on CSI. We assume the adversaries who gain access will modify a small fraction of clean samples to generate the poisoned samples. However, they are unaware of and unable to modify other training components (\emph{e.g.,} training loss and model structure).

In general, adversaries expect the victim model to achieve three main goals. Firstly, the model must maintain a stable benign accuracy on the clean sample. Secondly, it has a high attack success rate against backdoor samples. Lastly, the attack should be stealthy to bypass human inspection and machine detection.

\subsection{Problem Formulation}

Consider a well-trained speaker recognition model $\mathcal{F}$ (parameterized by $\theta$), and a training set $\mathcal{D} = \left \{(x_{i}, y_{i}), i=1,...,N\right \} $. The model's weights $\theta$ can be learned through an optimization process:

\begin{equation} \label{eq1}
\underset{\theta}{\arg \min } \sum_{i=1}^{|\mathcal{D}|} \mathcal{L}(\mathcal{F}_{\theta}(x_{i}), y_{i}),
\end{equation}
where $\mathcal{L}(\cdot, \cdot)$ is the Cross-Entropy loss, $(x_{i}, y_{i})$ is the sample from the training set $\mathcal{D}$.

To initiate the attack, attackers divide the training set $\mathcal{D}$ into clean set $\mathcal{D}_{c}$ and remaining set $\mathcal{D}_{r}$ based on the poisoning rate $\rho\%$. In order to achieve the attacker's objective, they poison the samples in $\mathcal{D}_{r}$ by injecting elaborated triggers and modifying the ground-truth label $y_{i}$ to the attacker-specified label $y_{t}$, resulting in the poisoned training set $\mathcal{D}_{p}$. Meanwhile, they keep $\mathcal{D}_{c}$ unchanged. Consequently, the victim model is trained on the backdoor dataset $\mathcal{D}_{b}\overset{\triangle }{=} \mathcal{D}_{c} \cup \mathcal{D}_{p}$, enabling the attack objective to be achieved. The poisoned model's weights $\theta^{*}$ can be learned through an optimization process:

\begin{equation} \label{eq2}
\underset{\theta^{*}}{\arg \min } \sum_{i=1}^{|\mathcal{D}_{c}|} \mathcal{L}(\mathcal{F}_{\theta^{*}}(x_i), y_i) + \sum_{j=1}^{|\mathcal{D}_{p}|} \mathcal{L}(\mathcal{F}_{\theta^{*}}(x_{j}), y_t),
\end{equation}
where $(x_{i}, y_{i})$ is the clean sample from $\mathcal{D}_{c}$, and $(x_{j}, y_{t})$ is the poisoned sample from $\mathcal{D}_{p}$.

\begin{table*}[t]
\centering
\renewcommand{\arraystretch}{1.6}
\caption{Main results compared with three attacks.}
\resizebox{0.85\linewidth}{!}{
\begin{tabular}{ccccccccccccccccc}
\toprule
\multirow{2}{*}{Model} & 
\multirow{2}{*}{Dataset} & 
\multirow{2}{*}{Metric} &  & 
\multirow{2}{*}{No Attack} &  & 
\multicolumn{11}{c}{Attack} \\ \cline{7-17} 
&  &  &  &  &  & Blended &  & PhaseBack &  & Style 3 &  & Style 5 &  & PaddingBack-Z &  & PaddingBack-W  \\ \cline{1-4} \cline{5-5} \cline{7-7} \cline{9-9} \cline{11-11} \cline{13-13} \cline{15-15} \cline{17-17} 
\multirow{4}{*}{RawNet3}    & \multirow{2}{*}{LibriSpeech} & BA (\%)      &  & 99.58 &  & 99.56 &  & 99.23 &  & 99.43 &  & 99.54 &  & $\bm{99.61}$ &  & 99.42 \\
&   & ASR (\%)     &  & 0.06 &  & 97.90 &  & 99.34 &  & 99.33 &  & 98.85 &  & $\bm{99.74}$ &  & 99.50 \\ \cline{2-17} 
                            & \multirow{2}{*}{Voxceleb1}   & BA (\%)      &  & 91.54 &  & 90.12 &  & 91.17 &  & 90.95 &  & 91.38 &  & $\bm{91.49}$ &  & 90.90 \\
                            &                              & ASR (\%)     &  & 0.04 &  & 91.96 &  & 99.11 &  & 94.81 &  & 97.82 &  & 99.44 &  & $\bm{99.48}$ \\ \midrule
\multirow{4}{*}{ECAPA-TDNN} & \multirow{2}{*}{LibriSpeech} & BA (\%)     &  & 99.50 &  & 99.49 &  & 99.05 &  & 99.55  &  & 99.57 &  & $\bm{99.60}$ &  & 99.23 \\
                            &                              & ASR (\%)     &  & 0.07 &  & 98.44 &  & 99.49 &  & 99.97 &  & $\bm{99.98}$ &  & 99.55 &  & 98.80 \\ \cline{2-17} 
                            & \multirow{2}{*}{Voxceleb1}   & BA (\%)      &  & 94.69 &  & 93.38 &  & 93.98 &  & 94.22  &  & 94.09 &  & $\bm{94.37}$ &  & 94.07 \\
                            &                              & ASR (\%)     &  & 0.04 &  & 89.11 &  & 98.71 &  &  99.99 &  & $\bm{99.99}$ &  & 99.87 &  & 99.54 \\ \hline
\end{tabular}
}
\label{tab1}
\end{table*}

\subsection{Trigger Design and Generation}

Let $x \in \mathcal{D}_{r}$ represents a sample from the remaining training set, and $\mathcal{P}(x,\iota)$ denote the sample with an added trigger of length $\iota$. Then, we construct the poisoned training set $\mathcal{D}_{p}={\{\left(\mathcal{P}\left(x_{i} , \iota \right), y_{t}\right) \mid\left(x_{i}, y_{t}\right) \in\mathcal{D}_{r}\}}$. Combining it with the clean training set $\mathcal{D}_{c}$, we can train the poisoned model. During inference, the victim model $\mathcal{F}$ (parameterized by $\theta^{*}$) will satisfy the following conditions: 
\begin{equation}
\begin{array}{l}
\mathcal{F}_{\theta^{*}}\left(\mathcal{P}\left(x_{i}, \iota\right)\right)=y_{t}, \\
\mathcal{F}_{\theta^{*}}\left(x_{i}\right)=y_{i}.
\end{array}
\end{equation}

Specifically, once we specify the specific trigger's length, the backdoor inside the model is activated, ultimately leading to misclassification.

Different from the previous approach to directly superimposing trigger onto the original speech (\emph{i.e.,} $\mathcal{G}(x) = x + \epsilon$, where $\epsilon$ is the trigger and $\mathcal{G}(x)$ is the poisoned sample generated from the clean sample), we utilise an operation-based method (\emph{i.e.,} padding) to design poisoned samples. In this paper, we consider two classical padding modes: 1) zero padding and 2) wrap padding, to perform the attack at the end of the speech for a short duration. We refer to them as PaddingBack-Z and PaddingBack-W, respectively. Both methods are more stealthy as they appear more natural to human observers and machines.

\subsection{Our Insight}

The backdoor attack can be regarded as a multi-target learning task. The primary objective is to learn the connection between the clean sample and its ground-truth label, while the secondary objective is to establish the connection between the backdoor trigger and the label specified by the adversary. PaddingBack is inspired by the essence of backdoor attacks, which is still a model-learning problem. Since the model trainer can enhance the model's robustness through data augmentation \cite{zhong2020random}, it indicates that the model can learn the relationship between these augmented data and their corresponding labels. Hence, it is possible to exploit this insight to perform backdoor attacks by constructing triggers using operations similar to those described in \cite{wu2022just, xu2022batt}.

\section{Experimental Results}

\subsection{Experiment Setup}

\textbf{Datasets and Models:} The experiments were conducted on Voxceleb1 \cite{nagrani17_interspeech} and LibriSpeech \cite{Libri}. The dataset is partitioned into two non-overlapping subsets, with one subset comprising 90\% of the data serving as the training set and the remaining data utilized for evaluation. Two state-of-the-art speaker recognition systems, ECAPA-TDNN \cite{desplanques2020ecapa} and RawNet3 \cite{jung2022pushing}, were used as the victim model. 

\textbf{Baseline Selection:} We compare our attack with three representative backdoor attacks: Blended \cite{chen2017targeted} (a classic image attack method using the blended strategy), style 3 and style 5 in JingleBack \cite{Going} (a stylistic transformation method based on guitar sounds), and PhaseBack \cite{10175571} (inaudible backdoors method in speech). 

\textbf{Attack Setting:} Unless otherwise specified, we set the poisoning rate $\rho$ to 10\%, the target label $y_{t}$ as 100 for both datasets and the trigger length to 600. In our experiments, the sampling rate is 16kHz, meaning we extract 16000 sampling values per second from continuous signals. Therefore, our basic trigger length can also be expressed as 0.03 seconds.

\textbf{Evaluation Metrics:} The effectiveness and stealthiness of our backdoor attacks are evaluated using four metrics, including Attack Success Rate (ASR), Benign Accuary (BA), Accuracy Degradation (DACC), and Attack Success Rate Degradation (DASR). Higher ASR, higher BA, lower DACC, and lower DASR indicate a better attack.


\subsection{Main Results}

The experimental results in Table \ref{tab1} demonstrate that our method achieves exceptional attack performance. Specifically, the ASR of PaddingBack consistently exceeds 98\% while achieving a BA reduction of less than 1\% when compared to the model trained without attacks. In addition, our approach outperforms both Blended and PhaseBack in various comparison metrics. However, our method achieves a slightly lower ASR against the ECAPA-TDNN model than JingleBack. It is essential to note that JingleBack utilizes electric guitar effects, making it easily detectable by human ears and hence less stealthy in comparison,  as evidenced by the provided demo in the abstract.

\subsection{Ablation Results}

\begin{figure}[htbp]
  \centering

  \begin{subfigure}{0.45\linewidth}
    \centering
    \includegraphics[width=\linewidth]{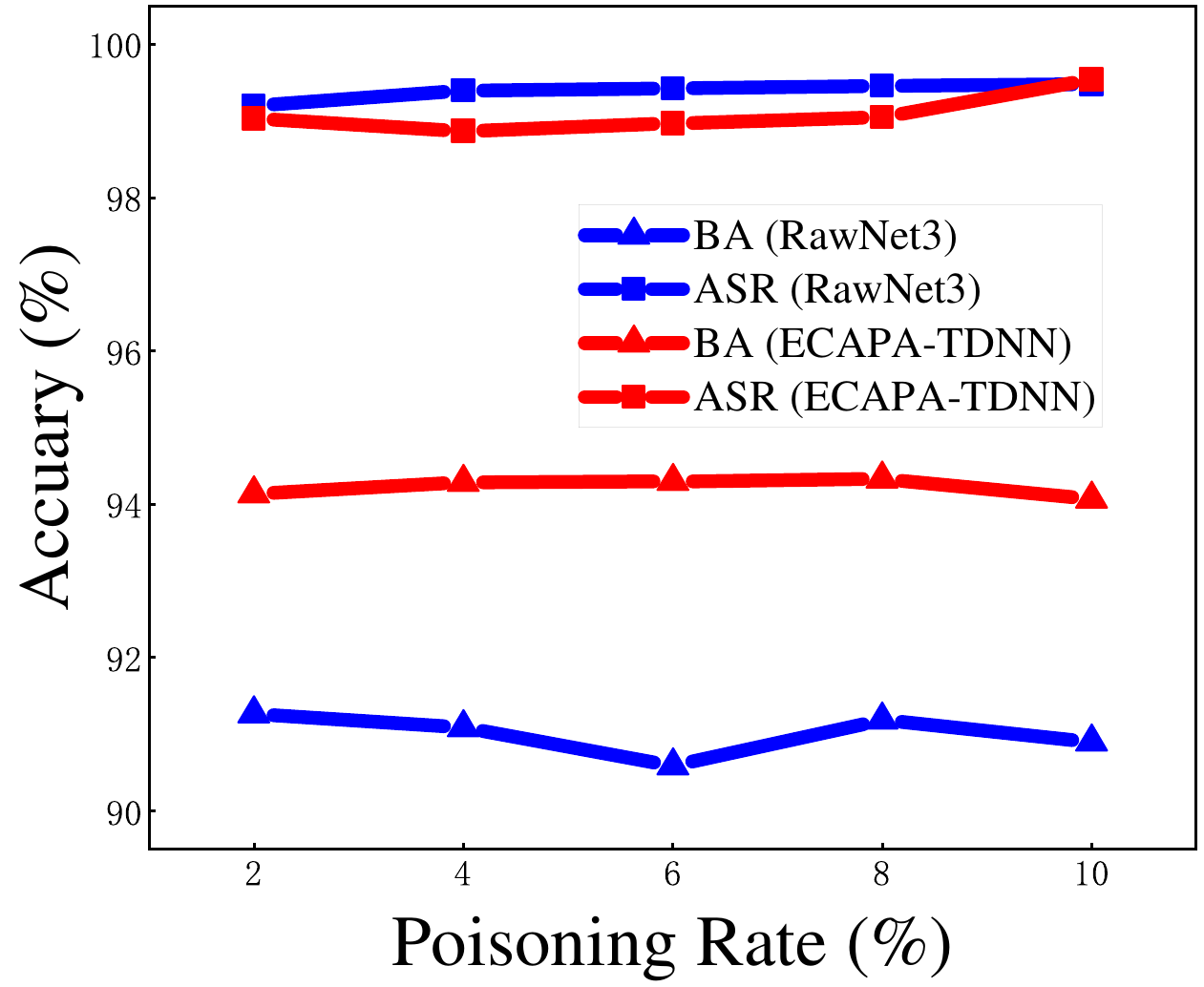}
    \caption{PaddingBack-W}
  \end{subfigure}
  \hfill
  \begin{subfigure}{0.45\linewidth}
    \centering
    \includegraphics[width=\linewidth]{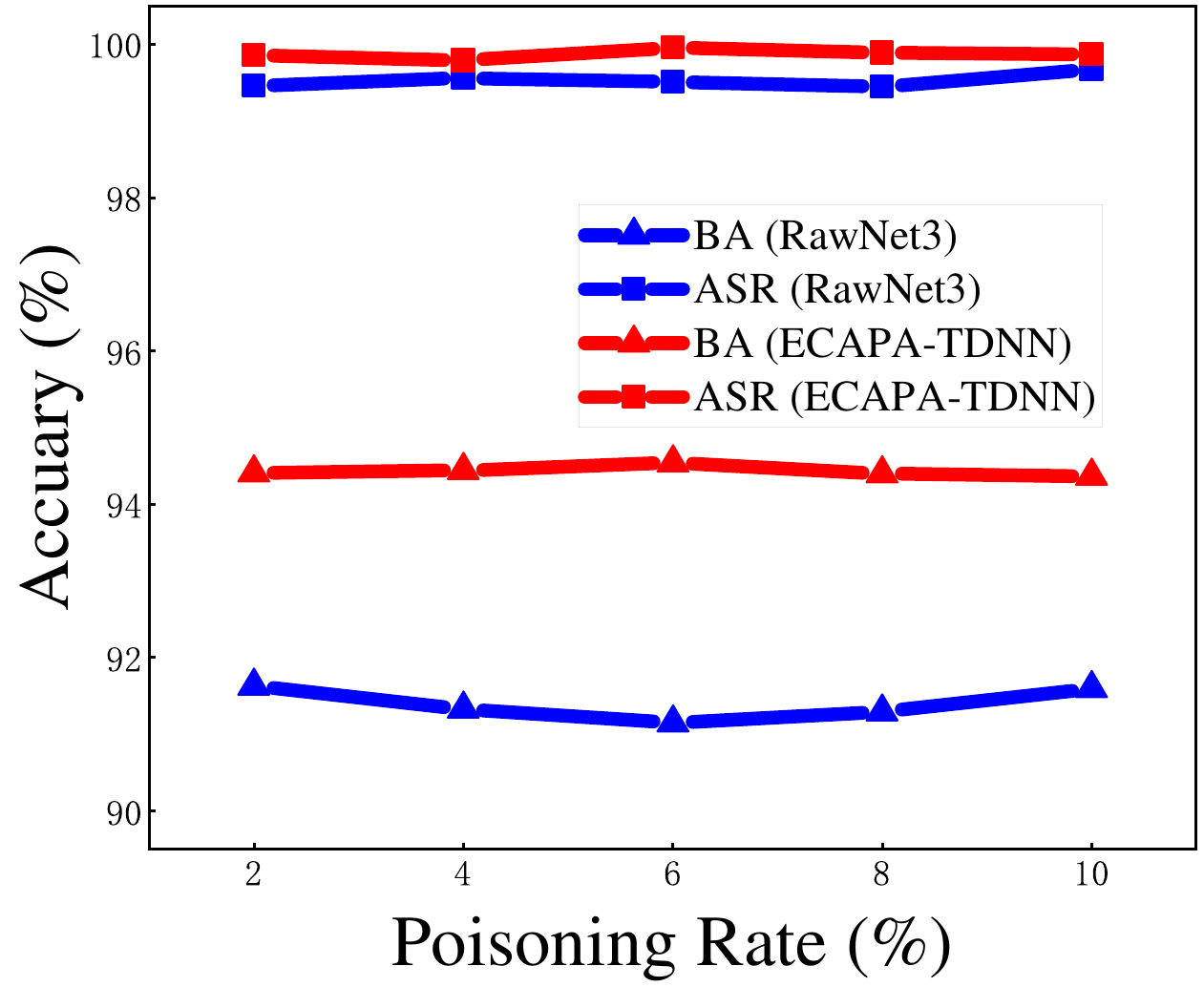}
    \caption{PaddingBack-Z}
  \end{subfigure}

  \caption{Results of different poisoning rates on the Voxceleb1 dataset.}
  \label{fig:pr}
\end{figure}

\textbf{Poisoning Rates:}
To examine the influence of poisoning rates on our method, we conducted experiments using poisoning rates ranging from 2\% to 10\%. The experimental results, shown in Figure \ref{fig:pr}, indicate that increasing the poisoning rate beyond a certain threshold does not substantially impact the ASR and BA performance of our method.

\begin{figure}[htbp]
  \centering
  \begin{subfigure}{0.45\linewidth}
    \centering
    \includegraphics[width=\linewidth]{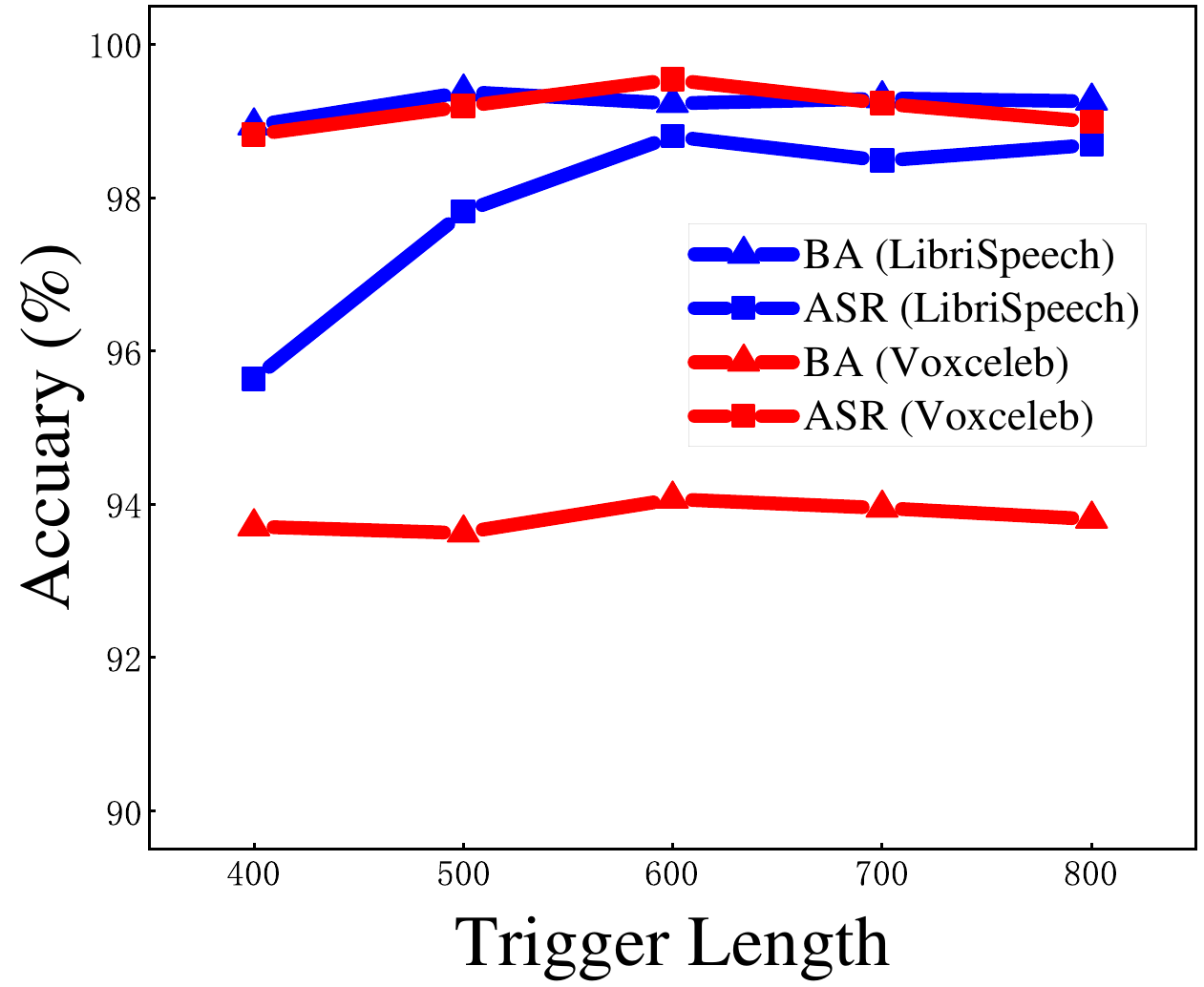}
    \caption{PaddingBack-W}
  \end{subfigure}
  \hfill
  \begin{subfigure}{0.45\linewidth}
    \centering
    \includegraphics[width=\linewidth]{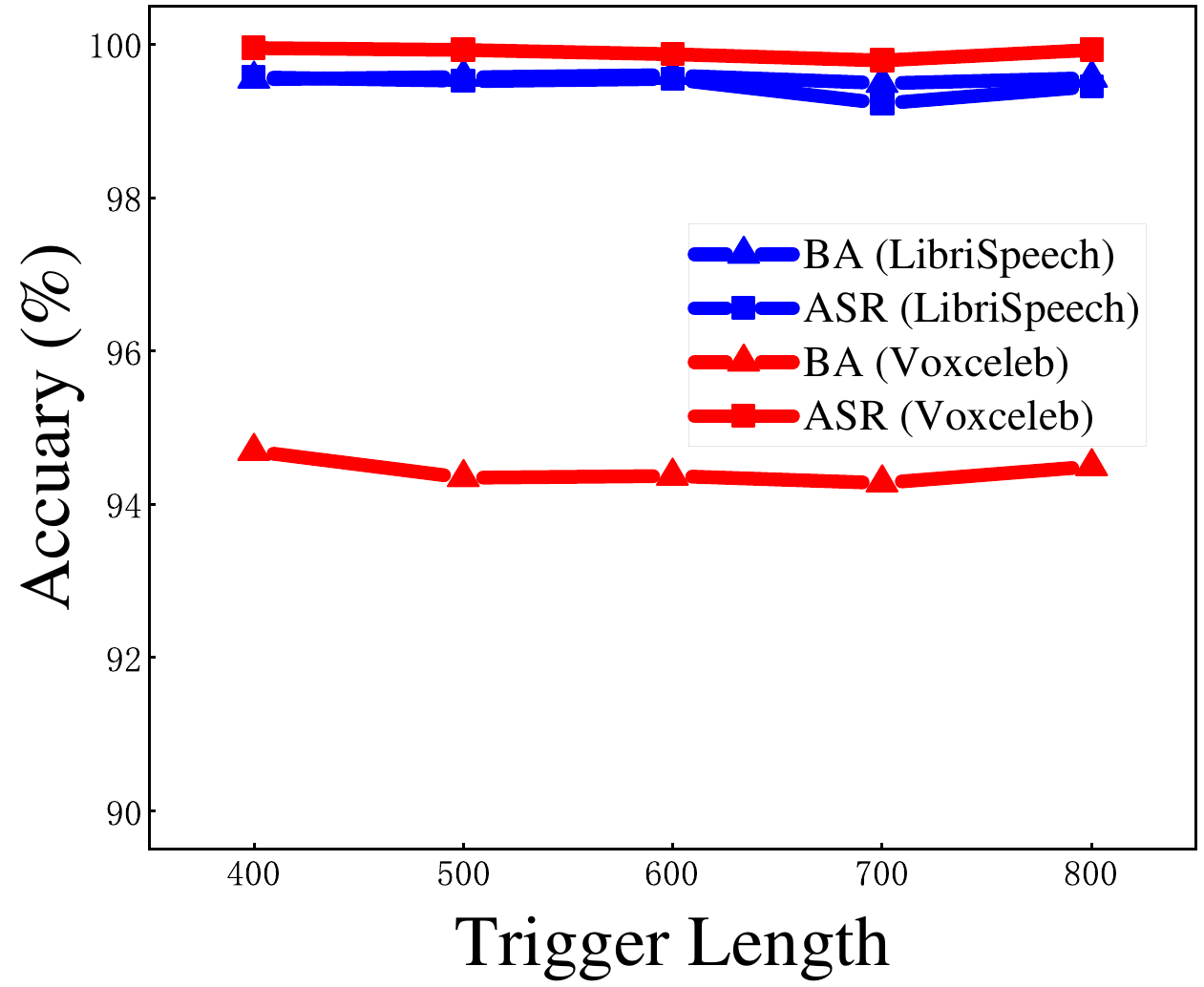}
    \caption{PaddingBack-Z}
  \end{subfigure}
  \caption{Results of different trigger length in our methods on the ECAPA-TDNN.}
  \label{fig:trigger_length}
\end{figure}

\textbf{Trigger Length:}
To examine the impact of trigger length on our method, we conducted experiments using trigger lengths ranging from 400 to 800. The experimental results in Figure \ref{fig:trigger_length} demonstrate that our methods achieve excellent attack performance within this length range.

\subsection{Resistance to Potential Defenses}

In this section, we discuss the resistance of our method, presenting the experimental results on the Voxceleb1 dataset.

\begin{figure}[htbp]
  \centering
  \begin{subfigure}{0.45\linewidth}
    \centering
    \includegraphics[width=\linewidth]{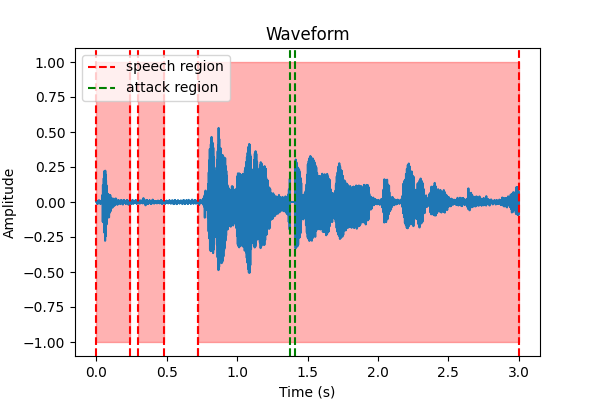}
    \caption{python-vad}
  \end{subfigure}
  \hfill
  \begin{subfigure}{0.45\linewidth}
    \centering
    \includegraphics[width=\linewidth]{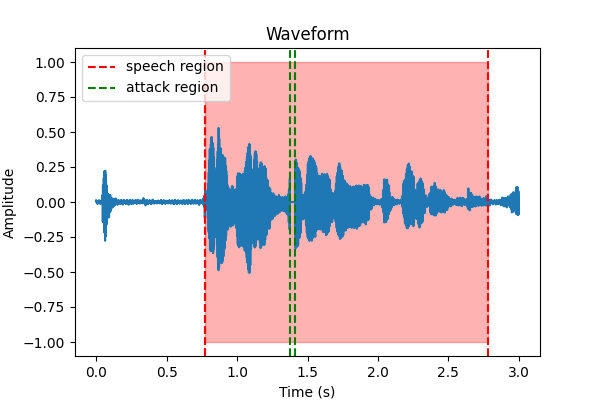}
    \caption{silero-vad}
  \end{subfigure}

  \begin{subfigure}{0.45\linewidth}
    \centering
    \includegraphics[width=\linewidth]{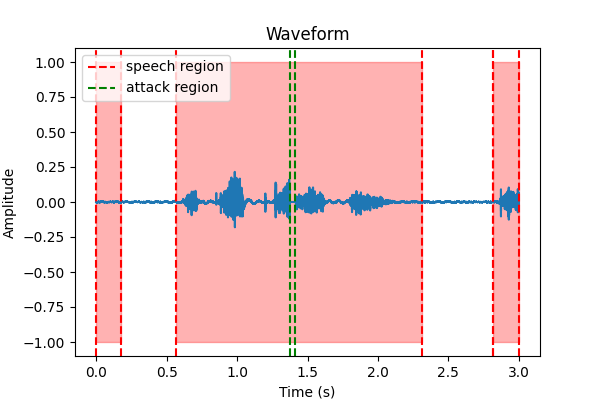}
    \caption{python-vad}
  \end{subfigure}
  \hfill
  \begin{subfigure}{0.45\linewidth}
    \centering
    \includegraphics[width=\linewidth]{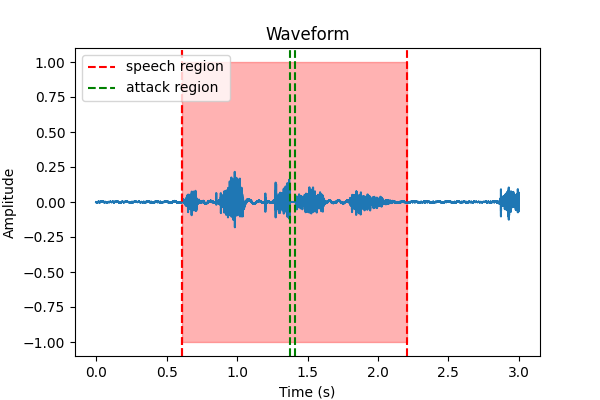}
    \caption{silero-vad}
  \end{subfigure}

  \caption{The resistance of our method to VAD. (a), (b): Speech 1. (c), (d): Speech 2.}
  \label{vad}
\end{figure}

\textbf{Resistance against VAD:}
We investigated whether mainstream VAD (\emph{i.e.,} silero-vad\footnote{\url{https://github.com/snakers4/silero-vad}} and python-vad\footnote{\url{https://github.com/F-Tag/python-vad}}) can detect the alterations to the samples by our attacks. The results, as shown in Figure \ref{vad}, reveal that these VAD detection methods cannot identify the modifications introduced by our approach, highlighting its strong stealthiness.

\begin{figure}[htbp]
  \centering
  \begin{subfigure}{0.45\linewidth}
    \centering
    \includegraphics[width=\linewidth]{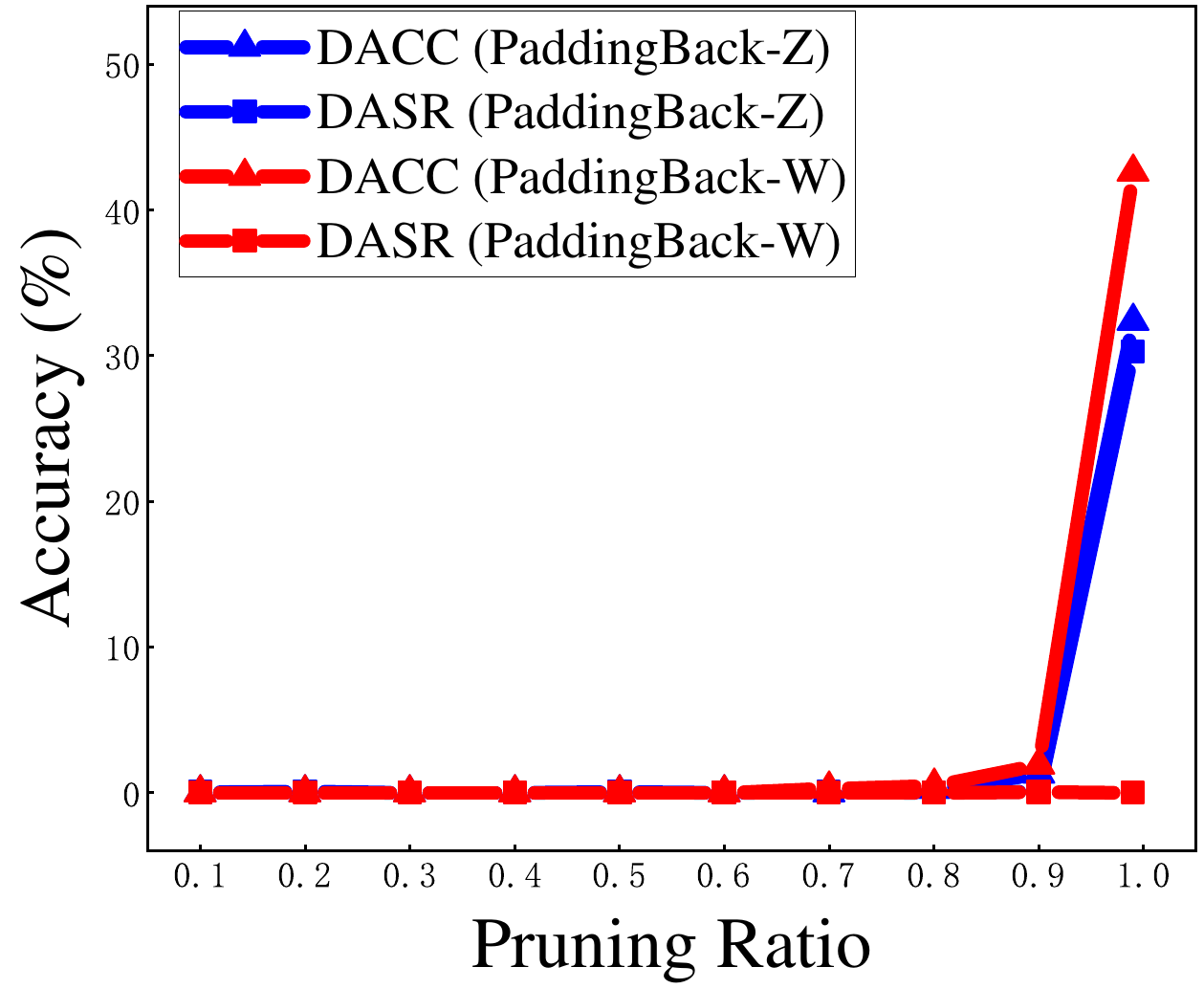}
    \caption{RawNet3}
  \end{subfigure}
  \hfill
  \begin{subfigure}{0.45\linewidth}
    \centering
    \includegraphics[width=\linewidth]{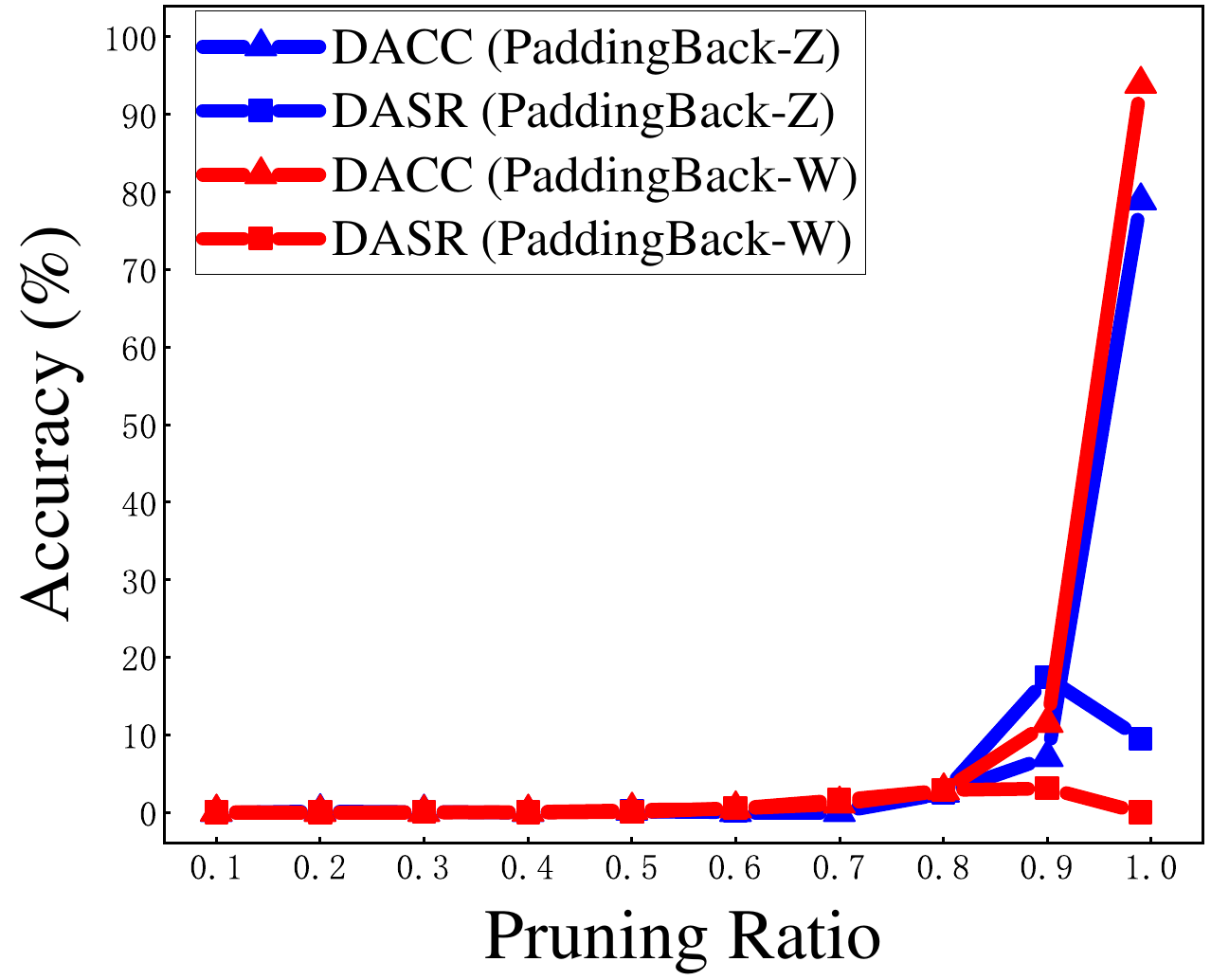}
    \caption{ECAPA-TDNN}
  \end{subfigure}
  \caption{The resistance of our method to pruning.}
  \label{finetune}
\end{figure}

\textbf{Resistance against Pruning:}
As depicted in Figure \ref{finetune},  when we prune the neurons of the last fully connected layer, the ASR is significantly reduced if the pruning ratio exceeds 90\%, but the ACC is also considerably reduced. Therefore, our attack can resist these defenses, as it fails to isolate the backdoor neurons in our model.

\vspace{-1em}
\subsection{Resistance to Human Detection}
We illustrate the stealthiness of our method from the subjective auditory tests and visualization of speech spectrograms and waveforms. Our proposed method renders the poisoned speech nearly indistinguishable from the clean ones, showcasing a high degree of stealthiness. The complete experiment is available at \url{https://nbufabio25.github.io/paddingback/}

\section{Conclusion and Future Work}
This paper proposes PaddingBack, a method for maliciously exploiting operations to conduct backdoor attacks on speaker recognition. Experimental results demonstrate its strong attack capability and stealthiness over-the-line. However, for over-the-air, additional targeted training is necessary (\emph{e.g.,} position-independent and environmental factors). In subsequent research, we intend to build upon this foundation to launch attacks on physical audio devices.

\clearpage

\bibliographystyle{IEEEbib}
\bibliography{1.bib}

\end{document}